\documentclass[sigconf]{acmart}

\usepackage{mathtools}
\usepackage{lipsum}
\usepackage{graphicx}
\usepackage{caption, subcaption}
\usepackage{url}
\usepackage{amsmath, amsfonts, amsthm, bm}
\usepackage{soul}
\usepackage{tabularx, multirow}
\usepackage{dsfont}
\usepackage[ruled,vlined]{algorithm2e}
\usepackage{booktabs}
\usepackage{xcolor, colortbl}
\usepackage{xspace}
\usepackage{resizegather}
\usepackage{makecell}
\usepackage[normalem]{ulem}
\usepackage{CJKutf8}
\usepackage{adjustbox}
\usepackage[utf8]{inputenc}
\usepackage[cjk]{kotex}
\usepackage{lipsum}
\usepackage{enumitem}

\usepackage[textsize=tiny]{todonotes}
\usepackage{marginnote}
 \let\marginpar\marginnote
\newcolumntype{W}{>{\centering\arraybackslash}m{0.08\linewidth}}
\newcolumntype{X}{>{\centering\arraybackslash}m{0.06\linewidth}}
\newcolumntype{Y}{>{\centering\arraybackslash}m{0.14\linewidth}}
\newcolumntype{Z}{>{\centering\arraybackslash}m{0.07\linewidth}}
\newcolumntype{P}{>{\centering\arraybackslash}m{0.19\linewidth}}
\newcolumntype{Q}{>{\raggedright\arraybackslash}m{0.28\linewidth}}
\newcolumntype{S}{>{\centering\arraybackslash}m{0.16\linewidth}}

\AtBeginDocument{%
  \providecommand\BibTeX{{%
    \normalfont B\kern-0.5em{\scshape i\kern-0.25em b}\kern-0.8em\TeX}}}

\setcopyright{acmcopyright}
\copyrightyear{2018}
\acmYear{2018}
\acmDOI{XXXXXXX.XXXXXXX}

\acmConference[Conference acronym 'XX]{Make sure to enter the correct
  conference title from your rights confirmation emai}{June 03--05,
  2018}{Woodstock, NY}
\acmPrice{15.00}
\acmISBN{978-1-4503-XXXX-X/18/06}

\begin{document}

\title{Towards Unified and Adaptive Cross-Domain Collaborative Filtering via Graph Signal Processing}


\author{Jeongeun Lee}
\affiliation{
    \institution{Yonsei University, South Korea}
    \country{}
}
\email{ljeadec31@yonsei.ac.kr}

\author{SeongKu Kang}
\affiliation{
    \institution{UIUC, United States}
    \country{}
}
\email{seongku@illinois.edu}

\author{Won-Yong Shin}
\affiliation{
    \institution{Yonsei University, South Korea}
    \country{}
}
\email{wy.shin@yonsei.ac.kr}

\author{Jeongwhan Choi}
\affiliation{
    \institution{Yonsei University, South Korea}
    \country{}
}
\email{jeongwhan.choi@yonsei.ac.kr}

\author{Noseong Park}
\affiliation{
    \institution{KAIST, South Korea}
    \country{}
}
\email{noseong@kaist.ac.kr}

\author{Dongha Lee}
\affiliation{
    \institution{Yonsei University, South Korea}
    \country{}
}
\authornote{Corresponding author.}
\email{donalee@yonsei.ac.kr}


\begin{abstract}
Collaborative Filtering (CF) is a foundational approach in recommender systems, but it struggles with challenges such as data sparsity and the cold-start problem. 
Cross-Domain Recommendation (CDR) has emerged as a promising solution by leveraging dense domains to improve recommendations in sparse target domains. 
However, existing CDR methods face significant limitations, including their reliance on overlapping users as a bridge between domains and their inability to address domain sensitivity, i.e., differences in user behaviors and characteristics across domains, effectively.
To overcome these limitations, 
we propose \proposed, a \textit{unified} and \textit{adaptive} CDR framework based on graph signal processing (GSP).
\proposed supports both intra-domain and inter-domain recommendations while adaptively controlling the influence of the source domain through a simple hyperparameter.
The framework constructs a cross-domain similarity graph by integrating target-only and source-bridged similarity graphs to capture both intra-domain and inter-domain relationships.
This graph is then processed through graph filtering techniques to propagate and enhance local signals. 
Finally, personalized graph signals are constructed, tailored separately for users in the source and target domains, enabling \proposed to function as a unified framework for CDR scenarios.
Extensive evaluation shows that \proposed outperforms state-of-the-art baselines across diverse cross-domain settings, with notable gains in low-overlap scenarios, underscoring its practicality for real-world applications.
\end{abstract}




\newcommand{\proposed}{\textsc{CGSP}\xspace}
\newcommand{\proposedone}{\textsc{CGSP}\textsubscript{\textsc{io}}\xspace}
\newcommand{\proposedtwo}{\textsc{CGSP}\textsubscript{\textsc{oa}}\xspace}
\newcommand{\proposedthree}{\textsc{CGSP}\textsubscript{\textsc{ua}}\xspace}

\newcommand{\proposedonef}{CGSP\textsubscript{\textsc{io}}$^\dagger$\xspace}
\newcommand{\proposedtwof}{{CGSP}\textsubscript{\textsc{oa}}$^\dagger$\xspace}
\newcommand{\proposedthreef}{{CGSP}\textsubscript{\textsc{ua}}$^\dagger$\xspace}

\newcommand{\proposedonev}{{CGSP}\textsubscript{\textsc{io}}$^*$\xspace}
\newcommand{\proposedtwov}{{CGSP}\textsubscript{\textsc{oa}}$^*$\xspace}
\newcommand{\proposedthreev}{{CGSP}\textsubscript{\textsc{ua}}$^*$\xspace}

\newcommand{\pgsp}{PGSP\xspace}
\newcommand{\gfcf}{GF-CF\xspace}
\newcommand{\lgcn}{LGCN-IDE\xspace}
\newcommand{\lightgcn}{LGCN\xspace}
\newcommand{\bpr}{BPR\xspace}

\newcommand{\dcdcsr}{DCDCSR\xspace}
\newcommand{\conet}{CoNet\xspace}

\newcommand{\emcdr}{EMCDR\xspace}
\newcommand{\sscdr}{SSCDR\xspace}
\newcommand{\ptupcdr}{PTUPCDR\xspace}
\newcommand{\unicdr}{UniCDR\xspace}


\newcommand{\veryshortarrow}[1][3pt]{\mathrel{%
\hbox{\rule[\dimexpr\fontdimen22\textfont2-.2pt\relax]{#1}{.4pt}}%
\mkern-4mu\hbox{\usefont{U}{lasy}{m}{n}\symbol{41}}}}
\makeatletter
\definecolor{Gray}{gray}{0.9}

\newcommand{\sinkhorn}{Sinkhorn\xspace}
\newcommand{\tsne}{TSNE\xspace}


\newcommand{\glove}{GloVe\xspace}
\newcommand{\bert}{BERT\xspace}
\newcommand{\sbert}{SBert\xspace}
\newcommand{\roberta}{RoBERTa\xspace}
\newcommand{\robertabase}{RoBERTa\textsubscript{base}\xspace}
\newcommand{\robertalarge}{RoBERTa\textsubscript{large}\xspace}

\newcommand{\douban}{Douban\xspace}
\newcommand{\amazon}{Amazon\xspace}

\newcommand{\reduce}[1]{\textls[-50]{#1}}
\newcommand{\smallsection}[1]{{\vspace{0.05in} \noindent \bf {#1.\hspace{5pt}}}}



\maketitle

\section{Introduction}
\label{sec:intro}
\begin{figure}[t]
    \centering    
    \includegraphics[width=\linewidth]{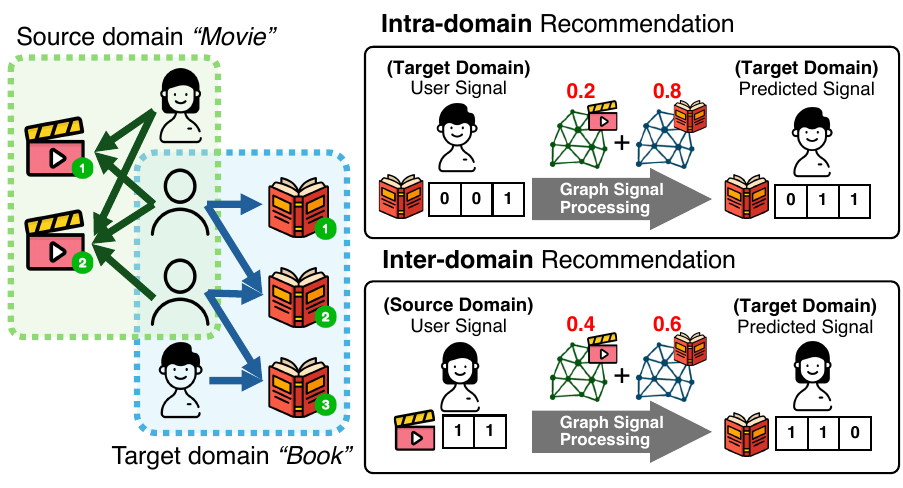} 
    \caption{Two cross-domain recommendation scenarios, with adaptive control of source and target domain importance.}
    \vspace{-0.3cm}
    \label{fig:intro}v
\end{figure} 

Collaborative filtering (CF) is a widely-adopted approach in recommender systems, which predicts user preference based on historical user-item interactions.~\cite{bokde2015matrix, he2017neural, rendle2012bpr}
Recently, graph convolutional networks (GCNs) have emerged as a powerful approach for CF, effectively leveraging user-item bipartite graphs to propagate information across nodes~\cite{he2020lightgcn, wang2019neural, berg2017graph}.
Notably, \lightgcn~\cite{he2020lightgcn} has shown superior performance by relying solely on graph structures, forgoing the need for nonlinear activations.
~\cite{xu2020neural, xu2021optimization} 
This result highlights the importance of fully exploiting the intrinsic structure of graphs while minimizing overall model complexity. 
Despite these advancements, CF methods remain constrained by persistent challenges such as data sparsity and the cold-start problem, which arise when users or items have limited historical interactions.


To overcome these challenges, cross-domain recommendation (CDR) offers a promising solution by leveraging dense source domains to enhance recommendation performance in sparse target domains.
CDR scenarios can be broadly categorized into two types: \textit{intra-domain} and \textit{inter-domain} recommendation. 
Intra-domain recommendation focuses on recommending items to users within the same domain, while leveraging shared user preferences or interaction patterns to mitigate the data sparsity problem of the target domain ~\cite{hu2018conet, li2020ddtcdr, cao2022disencdr, chen2018deep, huang2023cpr}.
On the other hand, inter-domain recommendation aims to recommend items to users in a different domain, where there exist discrepancies between the domains~\cite{zhu2021transfer, man2017cross, zhu2022personalized, kang2019semi}.
In this sense, transfer learning methods have been developed to map and align features between the source and target domains.

Despite their potential, existing CDR methods face two critical challenges. 
First, their performance heavily depends on the ratio of overlapping users—users present in both source and target domains. 
Since overlapping users serve as the primary bridge for transferring knowledge~\cite{man2017cross, chen2023cdradapterlearningadaptersdig}, a low overlap ratio significantly reduces the effectiveness of this transfer. 
Second, as highlighted in recent studies~\cite{cao2023towards, liu2020cross}, most methods overlook the fundamental differences in domain characteristics and user behaviors, referred to as domain sensitivity, which can lead to suboptimal performance. 
Additionally, most existing CDR methods are designed specifically for either intra-domain or inter-domain recommendation, limiting their generality and adaptability across different scenarios.

In this work, we explore Graph Signal Processing (GSP) as an innovative approach to address the challenges of CDR.
GSP emphasizes “\textbf{\textit{smoothness}},” a property that ensures similar users or items exhibit similar graph signals—a concept closely aligned with the core principles of CF~\cite{shen2021powerful}.
By representing users and items in the spectral domain through the Graph Fourier Transform (GFT), GSP uncovers latent patterns in complex graph structures.
Recent GSP-based CF methods, such as \gfcf~\cite{shen2021powerful} and \pgsp~\cite{liu2023personalized}, construct similarity graphs for users and items, smoothing them with graph filtering functions, including linear and ideal low-pass filters.
While \gfcf~\cite{shen2021powerful} focuses on item-based similarity graphs, PGSP~\cite{liu2023personalized} extends this by incorporating user similarities, demonstrating that richer graph structures can improve recommendation performance.
By leveraging interaction or similarity information as personalized graph signals, GSP refines these signals to enhance the prediction of user preferences.
However, these methods are primarily designed for single-domain scenarios, limiting their applicability to the complexities of CDR Scenarios.
A naive extension to merge source and target domain graphs based on overlapping users (i.e unified domain) often exacerbates sparsity, leading to performance degradation.
These limitations highlight the need for an effective integration strategy that constructs cross-domain similarities, enabling GSP to fully realize its potential in CDR applications.

This paper proposes \proposed, a novel \underline{\textbf{C}}DR framework based on \underline{\textbf{G}}raph \underline{\textbf{S}}ignal \underline{\textbf{P}}rocessing.
\proposed is designed to fuse preference knowledge from both source and target domains into a cross-domain similarity graph.
We propose various strategies for constructing the cross-domain similarity graph to support diverse CDR scenarios.
The core idea is to combine a target-only similarity graph, which captures relationships inherent within the target domain, with a source-bridged similarity graph, which incorporates inter-domain relationships.
To adaptively control the influence of the source domain depending on domain correlations, we introduce a simple yet effective hyperparameter.
Furthermore, \proposed derives personalized input signals separately for users in the source and target domains with their respective similarity information.
This enables \proposed to handle both intra-domain and inter-domain recommendation tasks within a unified framework.

To evaluate the practical applicability of \proposed, we conduct extensive experiments across diverse cross-domain settings, varying overlap ratios, sparsity levels, and domain characteristics. 
The results show taht \proposed consistently outperforms a wide range of baselines, including encoder-based models, GSP-based CF methods, and state-of-the-art CDR approaches, in both intra-domain and inter-domain scenarios. 
Notably, by effectively leveraging source-domain information, \proposed surpasses GSP-based CF models on single-domain or unified-domain graphs.
For the Amazon dataset known for its high sparsity, \proposed achieves a remarkable performance gain of over 20\% compared to the best baseline. 
Furthermore, \proposed demonstrates enhanced robustness by maintaining strong performance even under low user-overlap conditions.
It also offers significant time efficiency by avoiding parameter optimization, enabling faster execution compared to encoder-based baselines, making it well-suited for real-world CDR applications.
For reproducibility, our codes are publicly available at \href{https://github.com/ocryrtv/CGSP}{https://github.com/ocryrtv/CGSP}.

Our contributions can be summarized as follows:
\begin{itemize}
    \item We present \proposed, a unified GSP-based CDR framework that builds a cross-domain similarity graph with personalized signals for users in both source and target domains.
    \item To account for semantic correlations between domains, CGSP flexibly adjusts the influence of the source domain via a hyperparameter, thereby improving the effectiveness of cross-domain knowledge transfer. 
    \item \proposed shows notable performance compared to a wide range of baseline CF and CDR methods across diverse datasets, highlighting its effectiveness in both scenarios.
\end{itemize}

\section{Preliminary}
\label{sec:preliminary}

\subsection{Graph Signal Processing}
\label{subsec:gft}

\subsubsection{Notations for GSP} 
To deal with signals in a graph domain, a fundamental structure is an undirected simple graph $\mathcal{G}$ defined as an ordered pair ${\mathcal{G} = (\mathcal{V}, \mathcal{E})}$. 
Here, \({\mathcal{V}}\) represents a set of vertices \({\{v_1, v_2, \ldots, v_n\}}\), which correspond to entities (e.g., users or items), and \({\mathcal{E}}\) is a set of edges indicating relationships among the entities. 
The graph induces an adjacency matrix $\mathbf{A}$, where rows and columns respectively correspond to the nodes. 
An entry $\mathbf{A}_{ij}$ is set to 1 if there is an edge between node $i$ and $j$ in $\mathcal{G}$, and 0 otherwise. 

\subsubsection{Graph convolution} 

The concept of \textit{smoothness} in graphs can be mathematically represented as follows:
\begin{equation}
    S(x) = x^T \mathbf{L} x =  \sum_{i,j} \mathbf{A}_{i,j}(x_i - x_j)^2,
\end{equation}
where $\mathbf{L}=\mathbf{D}-\mathbf{A}$ is the Laplacian matrix.
Since $\mathbf{L}$ is real-valued and symmetric, its eigen-decomposition is expressed as $\mathbf{L} = \mathbf{U} \mathbf{\Lambda} \mathbf{U}^{\top}$, in which $\mathbf{U}$ is the matrix of eigenvectors and $\mathbf{\Lambda} = \text{diag}({\lambda_1}, {\lambda_2}, \ldots, {\lambda_n})$ is the diagonal matrix of eigenvalues with ${\lambda_1} \leq {\lambda_2} \leq \cdots \leq {\lambda_n}$. 

Graph Fourier Transform (GFT) decomposes graph signals into continuous frequency components. 
Formally, GFT of a signal $x$ defined on a graph $\mathcal{G}$ is given by $\hat{x} = \mathbf{U}^{\top} x$. 
Here, $\hat{x}$ represents the signal in the spectral domain, with each component indicating the signal's content at different frequencies defined by the eigenvalues in $\mathbf{\Lambda}$.
That is, GFT transforms the graph signal from its original spatial domain into the spectral domain, which represents frequency.

Note that eigenvectors associated with small eigenvalues enable a smoother division of the graph and capture similar feature information among the nodes. 
In contrast, eigenvectors with larger eigenvalues tend to distill distinct features between nodes. 
This disjunction highlights how GFT leverages the spectral properties of the graph to analyze the underlying structure of the data.
To be specific, given a Laplacian matrix $\mathbf{L}$, a graph filter can be defined by 
\begin{align}
H = \mathbf{U}\cdot\text{Diag}(h(\lambda_1 ), h(\lambda_2 ), \ldots, h(\lambda_n )) \cdot \mathbf{U}^\top,
\end{align}
where $h(\cdot)$ is the frequency response function. And with an input signal $x$, the filtered signal $y$ is obtained by 
\begin{align}
y = Hx = \mathbf{U} \cdot \text{Diag}(h(\lambda_1 ), h(\lambda_2 ), \ldots, h(\lambda_n )) \cdot \mathbf{U}^\top x.   
\end{align}
In the domain of GSP, graph signals are decomposed into diverse frequency components. 
Similar to traditional signal processing, low-frequency signals represent localized variations, marked by gradual changes across the graph, whereas high-frequency signals reflect global patterns, showing up through rapid changes across large parts of the graph. 

\subsubsection{Graph filtering methods}
There are several graph filtering methods employed for capturing different patterns (or frequency components) of the target graph.

\smallsection{Linear filter}
Given a normalized Laplacian matrix ${\bar{\mathbf{L}}}=\mathbf{I} - {\bar{\mathbf{A}}}$ and its eigenvalues $\lambda_i$, where $\mathbf{I}$ is the identity matrix and ${{\bar{\mathbf{A}}}}$ is a normalized adjacency matrix,
a first-order linear filter (also known as linear low-pass filter) locally smooths a graph as follows:
\begin{align}
h(\lambda) = 1 - \lambda.
\end{align}
%
The eigenvalues of ${\bar{\mathbf{A}}}$, denoted as $(\lambda_{\bar{\mathbf{A}}})_i$, are transformed by using
\begin{align}
(\lambda_{\bar{\mathbf{A}}})_i u_i = \bar{\mathbf{A}} u_i = (\mathbf{I} - \bar{\mathbf{L}}) u_i = u_i - \lambda_i u_i = (1 - \lambda_i) u_i. 
\end{align}

Thus, the eigenvalues $(\lambda_{\bar{\mathbf{A}}})_i$ of $\bar{\mathbf{A}}$ can be expressed as $1 - \lambda_i$. 
This relationship shows that the normalized adjacency matrix $\bar{\mathbf{A}}$ is effectively derived from applying the linear filter $H$ to the Laplacian matrix $\mathbf{L}$, indicated by $H={\bar{\mathbf{A}}}$.
This demonstrates the intrinsic connection between linear filtering of the Laplacian and the concept of similarity in the graph structure.

Applying a first-order linear filter is same as a one-layer spatial graph convolution. As both approaches rooted in neighborhood-based approaches~\cite{aiolli2013efficient, shen2021powerful}, the linear filter is simple yet effective in capturing local signals by exploring one-hop neighborhoods. 




\smallsection{Ideal low-pass filter}
An ideal low-pass filter passes signals with frequencies below a cutoff \(\lambda_c\) and attenuates those above it, effectively extracting a global representation of the signal.
Specifically, it is described by
\( h(\lambda) \) is defined as 1 for \( |\lambda| \leq \lambda_c \) and 0 for \( |\lambda| > \lambda_c \).



\subsection{GSP-based Collaborative Filtering}
\label{subsec:gsp-cf}
Recently, there have been several attempts to employ GSP for collaborative filtering (CF)~\cite{shen2021powerful, liu2023personalized}.
The main concept of GSP is \textit{smoothness}, which refers to the property of a signal on a graph where connected nodes (representing users or items in CF) have similar signal values. 
A smooth signal on the graph suggests that users and items connected by a short path within the graph are likely to interact with each other, indicating shared preferences. 
This concept aligns with the goal of CF, which aims to exploit similarities between users and items for recommendation.

Existing GSP-based CF methods involve three steps:
\begin{itemize}
    \item \textbf{Step 1.} Construct a similarity graph of users and items.
    \item \textbf{Step 2.} Apply a graph filter to the similarity graph.
    \item \textbf{Step 3.} Generate and filter personalized graph signals.
\end{itemize}

The first step is to build a similarity graph for users and items based on their interactions.
The graph encodes the underlying relationships among users, among items, and between users and items.
Next, the second step is to apply a graph filter. 
Existing methods~\cite{shen2021powerful, liu2023personalized, xia2022fire, xia2024hierarchical} apply a linear filter and an ideal low-pass filter, but our research adopts a focused approach by utilizing only a linear filter. 
The last step generates an input signal representing each user and then processes it through the filtered similarity graph;
this results in predicted (i.e., smoothed) scores for all items.

\begin{figure*}[thbp]
    \centering
    \includegraphics[width=\textwidth]{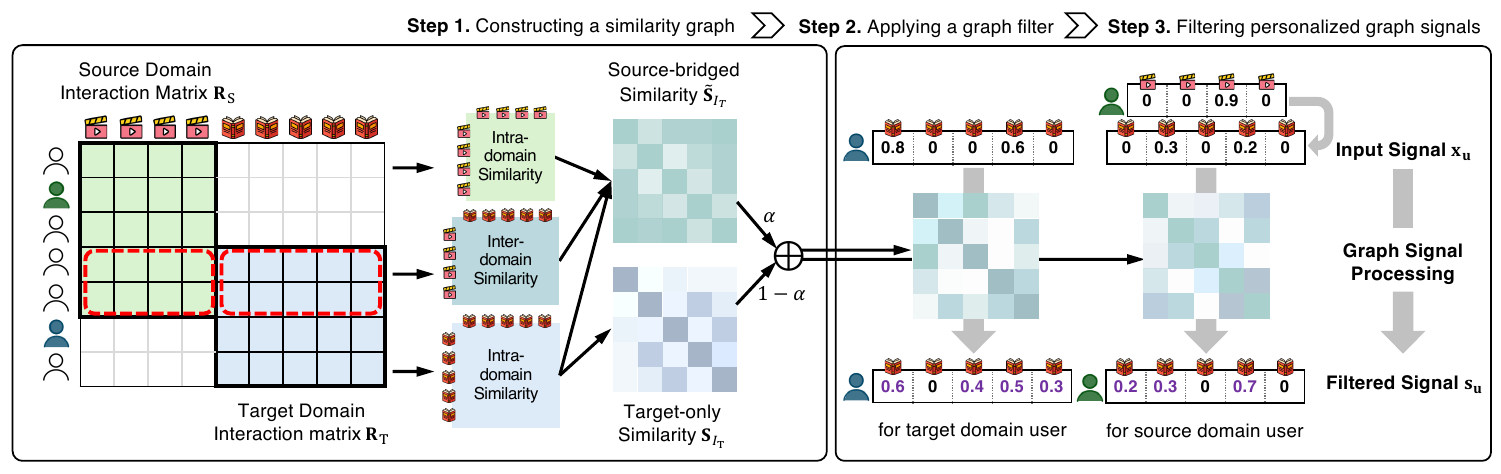} 
    \caption{The overall \proposed framework. It first constructs a cross-domain similarity graph $\mathbf{G}$ based on $\alpha$-weighted combination of source-bridged similarity $\widetilde{\mathbf{S}}$ and target-only similarity $\mathbf{S}$. Then, it processes personalized graph signals defined for source and target domain users, resulting in inter-domain and intra-domain recommendations, respectively. Note that the number of edges in the similarity graph is significantly smaller than  $|I_T|^2$, resulting in a low computational cost.}
    \label{fig:overview}
\end{figure*}

\section{Method}
\label{sec:method}
We present a unified CDR framework based on GSP, named as \proposed.
\proposed adopts the three basic steps of GSP-based CF, described in Section~\ref{subsec:gsp-cf}, tailoring it for two CDR scenarios: \textit{intra-domain} and \textit{inter-domain} recommendation.
\proposed effectively integrates preference knowledge from both source and target domains 
without encoder training, making CDR more robust to the ratio of overlapping users. 
\proposed also can readily adjust the contribution of the source domain information,
allowing for efficient knowledge utilization based on the degree of discrepancy between the domains.
The overview of our \proposed framework is illustrated in Figure~\ref{fig:overview}.


\subsection{Overview}
\label{subsec:overview} 
\subsubsection{Notations}
The source and target domains are represented as subscripts $S$ and $T$, respectively.
The sets of users and items in these domains are denoted as $\mathcal{U}_S$, $\mathcal{U}_T$ (for users), and $\mathcal{I}_S$, $\mathcal{I}_T$ (for items). 
We target the CDR setup where the source and target domains partially overlap only in users.
We define the set of overlapping users between the two domains as $\mathcal{U}_O = \mathcal{U}_S \cap \mathcal{U}_T$.
We define the normalized user-item interaction matrices for source and target domains, denoted by {$\mathbf{R}_S \in \mathbb{R}^{|\mathcal{U}_S| \times |\mathcal{I}_S|}$} and $\mathbf{R}_T \in \mathbb{R}^{|\mathcal{U}_T| \times |\mathcal{I}_T|}$, respectively. 
Each normalized matrix is obtained by $\mathbf{R}=\mathbf{D}_U^{-1/2} \widebar{\mathbf{R}} \mathbf{D}_I^{-1/2}$, where the original interaction matrix $\widebar{\mathbf{R}}$ has only binary values that indicate the absence or presence of interactions.\footnote{The normalized interaction matrix ensures equal impact from nodes of varying degrees, by penalizing the degree of highly active users and frequently rated items.}
For the overlapping users $\mathcal{U}_O$, we define $\mathbf{R}_{OS} \in \mathbb{R}^{|\mathcal{U}_O| \times |\mathcal{I}_S|}$ and $\mathbf{R}_{OT} \in \mathbb{R}^{|\mathcal{U}_O| \times |\mathcal{I}_T|}$ as their interaction matrix for each of the source and target domains.


The simplest similarity based on one-hop neighborhood is denoted as $\mathbf{S}$. 
For instance, within the target domain, we can define the basic similarity between users and items as $\mathbf{S}_{\mathcal{U}_T,\mathcal{I}_T} = \mathbf{R}_T$.
Based on this user-item similarity, the similarity between users and between items can be also specified as below: 
\begin{equation}
\label{eq:simgraph}
\begin{split}
\mathbf{S}_{\mathcal{U}_T} &= \mathbf{S}_{\mathcal{U}_T,\mathcal{I}_T} \mathbf{S}_{\mathcal{U}_T,\mathcal{I}_T}^\top = \mathbf{R}_T\mathbf{R}_T^\top, \\
   \mathbf{S}_{\mathcal{I}_T} &= \mathbf{S}_{\mathcal{U}_T,\mathcal{I}_T}^\top \mathbf{S}_{\mathcal{U}_T,\mathcal{I}_T} = \mathbf{R}_T^\top \mathbf{R}_T.
\end{split}
\end{equation}
Including all similarities of users and items~\cite{liu2023personalized}, the augmented similarity graph $\mathbf{S}_{\mathcal{U}_T \cup\mathcal{I}_T}\in\mathbb{R}^{(|\mathcal{U}_T|+|\mathcal{I}_T|)\times(|\mathcal{U}_T|+|\mathcal{I}_T|)}$ is defined by
\begin{equation}
    \mathbf{S}_{\mathcal{U}_T \cup\mathcal{I}_T} = 
\begin{bmatrix}
    \mathbf{S}_{\mathcal{U}_T} & \mathbf{S}_{\mathcal{U}_T, \mathcal{I}_T} \\
    (\mathbf{S}_{\mathcal{U}_T, \mathcal{I}_T})^\top & \mathbf{S}_{\mathcal{I}_T}
\end{bmatrix}.
\end{equation}


\subsubsection{GSP-based CDR framework}
The most straightforward solution for utilizing user-item interactions in both domains for GSP is to employ a unified graph that merges two graphs with overlapping users. 
This solution can be effective when the source and target domains share a sufficient number of users.
However, when there is minimal user overlap, the limited anchor information severely restricts effective knowledge transfer across domains;
this diminishes the benefits of utilizing the source domain. 
Therefore, we concentrate on developing a GSP-based CDR framework that can adeptly incorporate cross-domain information, enhancing the effectiveness of GSP even when direct user overlap is low.

Inspired by the concept of random walks \cite{perozzi2014deepwalk}, where nodes that are closely connected by a few steps are considered similar, we suggest that items from source and target domains could be seen as similar if there's a path connecting them through users common to both domains. 
This connection implies that these users have similar preferences in both domains, which could signify similarity between the items they interact with.
Building on this, our approach starts from capturing similarity between source domain items and target domain items and then explores the deeper connections between items (as well as between users) across different domains.

The first step of our \proposed framework is to construct a cross-domain similarity graph, denoted by $\mathbf{G}$;
this distills the compressed relationships between users and items across the two domains. 
It effectively addresses the data sparsity problem by constructing an interconnected graph that integrates direct similarities within the target domain with supportive similarities derived from the source domain. 
To effectively integrate source and target domain user-item information into a single similarity graph $\mathbf{G}$, we model two distinct graphs: 
i) the \textit{target-only} similarity graph $\mathbf{S}$ that encodes the information within the target domain, and ii) the \textit{source-bridged} similarity graph $\widetilde{\mathbf{S}}$ that incorporates additional information from the source domain. 
By linearly combining these graphs according to $\mathbf{G} = (1 - \alpha) \mathbf{S} + \alpha \widetilde{\mathbf{S}}$, \proposed captures signals across the domains as well as within the target domain. 
Note that the hyperparameter $\alpha$ optimizes the balance between two domains, controlling the extent to which source domain signals are utilized to enhance cross-domain recommendations.


Once the cross-domain similarity graph is constructed, a graph linear filter (see Section~\ref{subsec:gft}) is applied to the graph to strengthen the underlying connections. 
Then, an input signal for each user $x_u$ is generated to be filtered with the cross-domain similarity graph. 
Here, \proposed models more informative signals to encode more personalized preferences, instead of simply using a user's original interactions as the input signal.
By processing the personalized signals through the filtered graph, \proposed obtains the final prediction scores for each user.
In detail, for user $u$, the personalized signal $x_u$ is filtered by $s_u = x_u \mathbf{G}$, which serves as the final scores for item recommendation in the end. 

Based on the \proposed framework, in this work, we explore three different strategies to define and augment the cross-domain similarity graph, each of which is discussed in the following subsections.
\begin{itemize}
    \item Items-only similarity graph (\proposedone)
    \item Overlapping users-augmented similarity graph (\proposedtwo)
    \item Users-augmented similarity graph (\proposedthree)
\end{itemize}

\subsection{GSP on Items-Only Similarity Graph}
\label{subsec:cgsp-io}

\subsubsection{Cross-domain similarity graph construction}
\label{subsubsec:cgsp-io-g}
The first strategy is to construct a cross-domain similarity graph while focusing only on target domain items $\mathbf{G}_{\textsc{io}}\in\mathbb{R}^{|\mathcal{I}_T|\times|\mathcal{I}_T|}$.
The similarity graph is obtained by $\alpha$-weighted combination of the source-bridged similarity graph $\widetilde{\mathbf{S}}_{\mathcal{I}_T}$ and the target-only similarity graph ${\mathbf{S}}_{\mathcal{I}_T}$ (in Equation~\eqref{eq:simgraph}),
\begin{equation}
    \mathbf{G}_{\textsc{io}} = (1-\alpha) \mathbf{S}_{\mathcal{I}_T} + \alpha \widetilde{\mathbf{S}}_{\mathcal{I}_T}.
\end{equation}
The source-bridged item similarity graph is computed by incorporating information from the source domain to enhance the item similarity within the target domain.
Inter-domain item similarities are effectively inferred from interactions of the overlapping users.
\begin{equation}
\begin{split}
\widetilde{\mathbf{S}}_{\mathcal{I}_T} &= \mathbf{S}_{\mathcal{I}_T, \mathcal{I}_S} \mathbf{S}_{\mathcal{I}_S, \mathcal{I}_T} \mathbf{S}_{\mathcal{I}_T} = (\mathbf{R}_{OT}^\top \mathbf{R}_{OS}) (\mathbf{R}_{OS}^\top \mathbf{R}_{OT}) (\mathbf{R}_T^\top \mathbf{R}_T).
\end{split}
\end{equation}
With the help of source domain knowledge, it is capable of capturing more fine-grained relationships, thereby gaining more interconnected (or dense) representation of target domain items.


\subsubsection{Personalized graph signal filtering}
\label{subsubsec:cgsp-io-f}
To define personalized signals of each user for the cross-domain similarity graph, described in Section~\ref{subsubsec:cgsp-io-g}, we generate the signals based on the similarity between a user and the target domain items.
For intra-domain and inter-domain recommendation, we separately model the input signals for the target domain users $\mathbf{X}_T\in\mathbb{R}^{|\mathcal{U}_T|\times|\mathcal{I}_T|}$ and source domain users $\mathbf{X}_S\in\mathbb{R}^{|\mathcal{U}_S|\times|\mathcal{I}_T|}$, described as follows:
\begin{equation}
    \begin{split}
        \mathbf{X}_T &= \text{Signal}(\mathcal{U}_T \mapsto \mathcal{I}_T) = \mathbf{S}_{\mathcal{U}_T, \mathcal{I}_T} = \mathbf{R}_T, \\
        \mathbf{X}_S &= \text{Signal}(\mathcal{U}_S \mapsto \mathcal{I}_T)=\mathbf{S}_{\mathcal{U}_S, \mathcal{I}_S} \mathbf{S}_{\mathcal{I}_S, \mathcal{I}_T} = \mathbf{R}_S\mathbf{R}_{OS}^\top\mathbf{R}_{OT}.
    \end{split}
\end{equation}
Using the personalized signals above, the signal filtered on $\mathbf{G}_{\textsc{io}}$ can serve as the final scores for item recommendation.
That is, for each user $u$, the final score is obtained by $s_u = x_u \mathbf{G}_{\textsc{io}}\in\mathbb{R}^{|\mathcal{I}_T|}$, where $x_u$ denotes the input signal for the user $u$ from $\mathbf{X}$.

\subsection{GSP on Overlapping Users-Augmented Similarity Graph}
\label{subsec:cross-overlap}

\subsubsection{Cross-domain similarity graph construction}
\label{subsubsec:cgsp-ou-g}
The second strategy is to implement augmentation on items-only similarity graph $\mathbf{G}_{\textsc{io}}$ to include overlapping users $\mathcal{U}_O$ as well. The augmented similarity graph $\mathbf{G}_\textsc{oa} \in \mathbb{R} ^{(|\mathcal{U}_O|+|\mathcal{I}_T|) \times (|\mathcal{U}_O|+|\mathcal{I}_T|)}$ integrates source-bridged similarity $\widetilde{\mathbf{S}}_{\mathcal{U}_O\cup\mathcal{I}_T}$ and target-only similarity $\mathbf{S}_{\mathcal{U}_O\cup\mathcal{I}_T}$, both of which include all relationships between the overlapping users and target domain items.
\begin{equation}
\begin{split}
    &\mathbf{G}_\textsc{oa} = (1-\alpha)\mathbf{S}_{\mathcal{U}_O\cup\mathcal{I}_T} + \alpha\widetilde{\mathbf{S}}_{\mathcal{U}_O\cup\mathcal{I}_T} \\
    &= (1-\alpha) \begin{bmatrix}
    \mathbf{S}_{\mathcal{U}_O} & \mathbf{S}_{\mathcal{U}_O, \mathcal{I}_T} \\
    (\mathbf{S}_{\mathcal{U}_O, \mathcal{I}_T})^\top & \mathbf{S}_{\mathcal{I}_T}
    \end{bmatrix}
   + \alpha \begin{bmatrix}
    \widetilde{\mathbf{S}}_{\mathcal{U}_O} & \widetilde{\mathbf{S}}_{\mathcal{U}_O, \mathcal{I}_T} \\
    (\widetilde{\mathbf{S}}_{\mathcal{U}_O, \mathcal{I}_T})^\top & \widetilde{\mathbf{S}}_{\mathcal{I}_T}
    \end{bmatrix},
\end{split}
\raisetag{38pt}
\end{equation}
where the similarities of overlapping users in the target domain can be easily built by $\mathbf{S}_{\mathcal{U}_O} = \mathbf{R}_{OT}\mathbf{R}_{OT}^\top$ and $\mathbf{S}_{\mathcal{U}_O, \mathcal{I}_T} = \mathbf{R}_{OT} $, similar to Equation~\eqref{eq:simgraph}. 
The source-bridged similarity for overlapping users is attained by combining their similarities to both domains.
\begin{equation}
\begin{split}
\widetilde{\mathbf{S}}_{\mathcal{U}_O} &= \mathbf{S}_{\mathcal{U}_O, \mathcal{I}_T} \mathbf{S}_{\mathcal{I}_T, \mathcal{I}_S} 
\mathbf{S}_{\mathcal{I}_S} 
\mathbf{S}_{\mathcal{I}_S, \mathcal{I}_T} \mathbf{S}_{\mathcal{I}_T, \mathcal{U}_O} \\
&= \mathbf{R}_{OT}(\mathbf{R}_{OT}^\top \mathbf{R}_{OS}) (\mathbf{R}_S^\top \mathbf{R}_S) (\mathbf{R}_{OS}^\top \mathbf{R}_{OT})\mathbf{R}_{OT}^\top ,\\
\widetilde{\mathbf{S}}_{\mathcal{U}_O, \mathcal{I}_T} &= \mathbf{S}_{\mathcal{U}_O, \mathcal{I}_S} \mathbf{S}_{\mathcal{I}_S} \mathbf{S}_{\mathcal{I}_S, \mathcal{I}_T} \mathbf{S}_{\mathcal{I}_T, \mathcal{I}_S}  \\ 
&= \mathbf{R}_{OS}(\mathbf{R}_S^\top \mathbf{R}_S) (\mathbf{R}_{OS}^\top \mathbf{R}_{OT}) (\mathbf{R}_T^\top \mathbf{R}_T).
\end{split}
\end{equation}


\subsubsection{Personalized graph signal filtering}
\label{subsubsec:cgsp-ou-f}
Since the dimension of cross-domain similarity graph has been extended to ${|\mathcal{U}_O|+|\mathcal{I}_T|}$,
the input signal should also be extended to include similarity with the overlapping users, resulting in $\mathbf{X}_T\in\mathbb{R}^{|\mathcal{U}_T|\times(|\mathcal{U}_O|+|\mathcal{I}_T|)}$ and $\mathbf{X}_S\in\mathbb{R}^{|\mathcal{U}_S|\times(|\mathcal{U}_O|+|\mathcal{I}_T|)}$. Thus, input signals can be built by simply concatenating the similarities for overlapping users with those for target domain items: 
\begin{equation}
\begin{split}
\mathbf{X}_T &= 
    \text{Signal}(\mathcal{U}_T \mapsto \mathcal{U}_O \cup \mathcal{I}_T) \\
    &= \begin{bmatrix}
    \mathbf{S}_{\mathcal{U}_T, \mathcal{U}_O} & \mathbf{S}_{\mathcal{U}_T, \mathcal{I}_T}
\end{bmatrix}
    = \begin{bmatrix}
    \mathbf{R}_T\mathbf{R}_{OT}^\top & \mathbf{R}_T
\end{bmatrix}, \\
\mathbf{X}_S &= 
    \text{Signal}(\mathcal{U}_S \mapsto \mathcal{U}_O \cup \mathcal{I}_T) \\
    &= \begin{bmatrix}
    \mathbf{S}_{\mathcal{U}_S, \mathcal{U}_O} & \mathbf{S}_{\mathcal{U}_S, \mathcal{I}_T}
\end{bmatrix}
= \begin{bmatrix}
    \mathbf{R}_S\mathbf{R}_{OS}^\top & \mathbf{R}_S \mathbf{R}_{OS}^\top \mathbf{R}_{OT}
\end{bmatrix}.  
\end{split}
\end{equation}
Given the extended input signal $\mathbf{X}_T$, $\mathbf{X}_S$, and the cross-domain similarity graph $\mathbf{G}_{\textsc{oa}}$, a post-processing step is necessary for score prediction. 
This involves $s_u = (x_u \mathbf{G}_{\textsc{oa}})_{[-\text{\textit{n\_items}}:]}$, where \textit{n\_items} refers to the number of target domain items, which extracts each user's personalized preferences specifically towards the items.

\subsection{GSP on Users-Augmented Similarity Graph}
\label{subsec:cross-user}

\subsubsection{Cross-domain similarity graph construction}
For the last strategy, we extend the cross-domain similarity to include all users and items in the target domain. 
Being consistent with previous approaches in Sections~\ref{subsec:cgsp-io} and \ref{subsec:cross-overlap}, the source-bridged similarity $\widetilde{\mathbf{S}}_{\mathcal{U}_T\cup\mathcal{I}_T}$ is combined with target-only similarity $\mathbf{S}_{\mathcal{U}_T \cup\mathcal{I}_T}$, which results in the cross-domain similarity  $\mathbf{G}_\textsc{ua} \in \mathbb{R} ^{(|\mathcal{U}_T|+|\mathcal{I}_T|)\times(|\mathcal{U}_T|+|\mathcal{I}_T|)}$:
\begin{equation}
\begin{split}
    &\mathbf{G}_{\textsc{ua}} = (1-\alpha)\mathbf{S}_{\mathcal{U}_T \cup\mathcal{I}_T} + \alpha\widetilde{\mathbf{S}}_{\mathcal{U}_T\cup\mathcal{I}_T} \\
    &= (1-\alpha) \begin{bmatrix}
    \mathbf{S}_{\mathcal{U}_T} & \mathbf{S}_{\mathcal{U}_T, \mathcal{I}_T} \\
    (\mathbf{S}_{\mathcal{U}_T, \mathcal{I}_T})^\top & \mathbf{S}_{\mathcal{I}_T}
    \end{bmatrix}
   + \alpha \begin{bmatrix}
    \widetilde{\mathbf{S}}_{\mathcal{U}_T} & \widetilde{\mathbf{S}}_{\mathcal{U}_T, \mathcal{I}_T} \\
    (\widetilde{\mathbf{S}}_{\mathcal{U}_T, \mathcal{I}_T})^\top & \widetilde{\mathbf{S}}_{\mathcal{I}_T}
    \end{bmatrix}.
\end{split}
\raisetag{38pt}
\end{equation}
Note that the source-bridged similarity for target domain users $\widetilde{\mathbf{S}}_{\mathcal{U}_T}$ can be acquired by linking their similarities to target domain items $\mathbf{S}_{\mathcal{U}_T, \mathcal{I}_T}$ with inter-domain item similarities $\mathbf{S}_{\mathcal{I}_T, \mathcal{I}_S}$.
\begin{equation}
\begin{split}
\widetilde{\mathbf{S}}_{\mathcal{U}_T} &= \mathbf{S}_{\mathcal{U}_T, \mathcal{I}_T} \mathbf{S}_{\mathcal{I}_T, \mathcal{I}_S} 
\mathbf{S}_{\mathcal{I}_S} 
\mathbf{S}_{\mathcal{I}_S, \mathcal{I}_T} \mathbf{S}_{\mathcal{I}_T, \mathcal{U}_T} \\
&= \mathbf{R}_T(\mathbf{R}_{OT}^\top \mathbf{R}_{OS}) (\mathbf{R}_S^\top \mathbf{R}_S) (\mathbf{R}_{OS}^\top \mathbf{R}_{OT})\mathbf{R}_T^\top ,\\
\widetilde{\mathbf{S}}_{\mathcal{U}_T, \mathcal{I}_T} &= \mathbf{S}_{\mathcal{U}_T, \mathcal{I}_T} \mathbf{S}_{\mathcal{I}_T, \mathcal{I}_S} \mathbf{S}_{\mathcal{I}_S} 
\mathbf{S}_{\mathcal{I}_S, \mathcal{I}_T} \mathbf{S}_{\mathcal{I}_T} \\
&= \mathbf{R}_T(\mathbf{R}_{OT}^T \mathbf{R}_{OS})(\mathbf{R}_{S}^\top\mathbf{R}_{S})(\mathbf{R}_{OS}^\top \mathbf{R}_{OT})(\mathbf{R}_T^\top \mathbf{R}_T).
\end{split}
\end{equation}

\subsubsection{Personalized graph signal filtering}
To make the signals align with the augmented cross-domain similarity graph $\mathbf{G}_{\textsc{ua}}$, the personalized input signals broaden beyond the similarity with target domain items to encompass the similarity with all users in the target domain. 
They are built by concatenating the similarities with target domain users to those with target domain items.
\begin{equation}
\begin{split}
\mathbf{X}_T &= \text{Signal}(\mathcal{U}_T \mapsto \mathcal{U}_T\cup\mathcal{I}_T) \\
&= \begin{bmatrix}
       \mathbf{S}_{\mathcal{U}_T} & \mathbf{S}_{\mathcal{U}_T, \mathcal{I}_T}
   \end{bmatrix} 
   = \begin{bmatrix}
       \mathbf{R}_T\mathbf{R}_T^\top & \mathbf{R}_T
   \end{bmatrix} ,\\
\mathbf{X}_S &= \text{Signal}(\mathcal{U}_S \mapsto \mathcal{U}_T\cup\mathcal{I}_T) \\
&= \begin{bmatrix}
    \mathbf{S}_{\mathcal{U}_S, \mathcal{U}_T} & \mathbf{S}_{\mathcal{U}_S, \mathcal{I}_T}
\end{bmatrix}
= \begin{bmatrix}
    \mathbf{R}_S\mathbf{R}_{OS}^\top \mathbf{R}_{OT}\mathbf{R}_T^\top & \mathbf{R}_S\mathbf{R}_{OS}^\top \mathbf{R}_{OT}
\end{bmatrix}.
\end{split}
\raisetag{38pt}
\end{equation}
To extract individual user preferences for target domain items from the augmented graph $\mathbf{G}_{\textsc{ua}}$, it has to be post-processed to obtain the final predicted score. 
Using the personalized signal $ x_u \in \mathbb{R} ^{(|\mathcal{U}_T|+|\mathcal{I}_T|)}$, the final score is obtained by $s_u = (x_u \mathbf{G}_{\textsc{ua}})_{[-\text{\textit{n\_items}}:]}$.


\subsection{Strategy Comparison}
The aforementioned three strategies aim to construct a cross-domain similarity graph $\mathbf{G}$ incorporating the target-only similarity $\mathcal{S}$ and the source-bridged similarity $\widetilde{\mathcal{S}}$. 
All the strategies basically include the enriched item-item similarity, denoted by $\mathbf{G}_{\textsc{io}}$. 
However, \proposedone relies solely on item similarities between source and target domains, lacking in fully capturing the relationships in the target domain;
this highlights the importance of leveraging user-focused similarities. 
In this sense, \proposedtwo emphasizes the role of the overlapping users as the key bridge linking the two domains, while reflecting the inherent similarity within each domain.
Expanding further, \proposedthree includes all users in the target domain, considering the user similarity between source and target domains. 
The inclusion of users enhances the richness of the graph, facilitating more effective processing of personalized graph signals.

\section{Experiments}
\label{sec:exp}
In this section, we present the experimental results to answer the following research questions:
\begin{itemize}
    \item \textbf{RQ1}: How effective is \proposed to utilize source domain information with graph signal processing?
    \item \textbf{RQ2}: How effective is \proposed for the cold-start problem?
    \item \textbf{RQ3}: How effective is \proposed when the ratio of overlapping users is low between the source and target domains?
    \item \textbf{RQ4}: How effective is $\alpha$ to handle domain discrepancy?
    \item \textbf{RQ5}: How efficient is \proposed in terms of execution time?
\end{itemize}

\subsection{Experimental Settings}
\label{subsec:expset}

\subsubsection{Datasets and domain setup}

We employ two popular multi-domain datasets: \textbf{\douban} and \textbf{\amazon} \cite{zhu2020graphical, zhu2019dtcdr, kang2019semi}.
\douban is derived from a prominent Chinese social networking platform that allows users to review and rate movies, books, and music. 
\amazon is a comprehensive collection of user reviews and ratings from the Amazon e-commerce platform, encompassing a wide variety of products. 
Following the domain setup commonly used in recent studies~\cite{Zhu2018ADF, zhu2022personalized, cao2023towards}, we consider Movie as the source domain and Music and Book as the target domains for \douban. 
Additionally, for \amazon, we select Sports and Clothes as the source-target domain pairs~\cite{cao2023towards}.
It is worth noting that the two datasets have distinct characteristics. 
The \amazon dataset is highly sparse, which makes recommendation tasks challenging.
On the other hand, the \douban dataset has a very high ratio of overlapping users\footnote{This ratio is calculated as the proportion of target-domain users who are also present in the source domain.}, almost reaching 1;
this indicates that most of the users in the Music or Book domains are active in the Movie domain as well. 

The dataset statistics and domain setups are in Table~\ref{table:dataset} and Table~\ref{table:task}.
Our experiments encompass various cross-domain scenarios with different domain pairs, overlap ratios, and sparsity levels.

\subsubsection{Evaluation setup}
Our experiments include both intra-domain and inter-domain recommendation scenarios. 

We focus on the top-$N$ recommendation task for implicit feedback, adopting two commonly used ranking metrics: Recall@$N$ and NDCG@$N$ \cite{kang2019semi, cao2023towards}. 
Following our main baselines~\cite{he2020lightgcn, shen2021powerful, liu2023personalized}, we set $N$ to 20.
For a more~thorough~assessment, we choose a full-rank evaluation over a sampling-based evaluation, such as the leave-one-out protocol.


\begin{table}[t]
\small
\caption{Statistics of the datasets.}
\renewcommand{\arraystretch}{0.8}
\centering
\setlength{\tabcolsep}{3pt}
\begin{tabular}{c ccccc}
\toprule
\textbf{Datasets} & \textbf{Domain} & \textbf{\#Users} & \textbf{\#Items} & \textbf{Sparsity (\%)} \\
\midrule
\multirow{3}{*}{\textbf{\douban}} & Movie & \ \ 2,712 & 34,893 & 98.65 \\
& Music & \ \ 1,820 & 79,878 & 99.88 \\
& Book & \ \ 2,212 & 95,875 & 99.89 \\
\midrule
\multirow{4}{*}{\textbf{\amazon}} & Movie & 46,415 & 51,925 & 99.95 \\
& Music & 39,074 & 67,677 & 99.97 \\
& Sports & 19,119 & 41,082 & 99.97 \\
& Clothes & 41,475 & 80,249 & 99.99 \\
\bottomrule
\end{tabular}
\label{table:dataset}
\end{table}
\begin{table}[t]
\centering
\footnotesize
\renewcommand{\arraystretch}{0.75}
\setlength{\tabcolsep}{1.2mm}
\caption{Intra-domain and inter-domain CDR scenarios with different ratios of overlapping users.}
\resizebox{0.9\linewidth}{!}{
\begin{tabular}{c cccc}
\toprule
\textbf{Scenario} & \textbf{Dataset} & \textbf{Source}$\rightarrow$\textbf{Target} & \textbf{Overlap Ratio} \\
\midrule
\multirow{3}{*}{Intra-domain} & \douban & Movie$\rightarrow$Music & 1.000 \\
 & \douban & Movie$\rightarrow$Book & 0.999 \\
 & \amazon & Movie$\rightarrow$Music & 0.181 \\
\midrule
\multirow{3}{*}{Inter-domain} & \douban & Movie$\rightarrow$Music & 0.995 \\
 & \amazon & Movie$\rightarrow$Music & 0.099 \\
 & \amazon & Sports$\rightarrow$Clothes & 0.041 \\
\bottomrule
\end{tabular}
}
\label{table:task}
\end{table}



\begin{table*}[t!]
\small
\centering
\caption{Performance for top-$K$ recommendation to target domain users (\textit{intra-domain}; Left) and source domain users (\textit{inter-domain}; Right). 
$\dagger$ and $*$ report test accuracies with the fixed $\alpha=0.85$ and the optimal $\alpha$ showing the highest accuracy on the validation set, respectively. The best and second-best results are highlighted in bold and underlined.}
\label{tbl:intra-inter-domain}
\renewcommand{\arraystretch}{0.9}
\resizebox{0.99\linewidth}{!}{%
\begin{tabular}{ccXXXXXXXXXXXX}
    \toprule
    & \multirow{4.5}{*}{\textbf{Method}} & \multicolumn{6}{c}{\small \textbf{Intra-domain Recommendation}} & \multicolumn{6}{c}{\small \textbf{Inter-domain Recommendation}} \\
    \cmidrule(lr){3-8}\cmidrule(lr){9-14}
    
    & & \multicolumn{2}{Y}{\small \textbf{\douban Movie$\rightarrow$Music}}
    & \multicolumn{2}{Y}{\small \textbf{\douban Movie$\rightarrow$Book}} 
    & \multicolumn{2}{Y}{\small \textbf{\amazon Movie$\rightarrow$Music}} 
    & \multicolumn{2}{Y}{\small \textbf{\douban Movie$\rightarrow$Music}} 
    & \multicolumn{2}{Y}{\small \textbf{\amazon Movie$\rightarrow$Music}} 
    & \multicolumn{2}{Y}{\small \textbf{\amazon Sports$\rightarrow$Clothes}}
    \\

    & & \footnotesize Recall@20 & \footnotesize NDCG@20 
    & \footnotesize Recall@20 & \footnotesize NDCG@20 
    & \footnotesize Recall@20 & \footnotesize NDCG@20
    & \footnotesize Recall@20 & \footnotesize NDCG@20 
    & \footnotesize Recall@20 & \footnotesize NDCG@20 
    & \footnotesize Recall@20 & \footnotesize NDCG@20 \\
    \midrule


    \multirow{5.5}{*}{\rotatebox[origin=c]{90}{\footnotesize Single domain}}
     & \bpr
	& 0.0033 & 0.0019 
	& 0.0031 & 0.0011  
	& 0.0011 & 0.0015
        & - & - 
        & - & - 
        & - & - \\
 
    & \lightgcn
    & 0.0270 & 0.0200 
    & 0.0424 & 0.0282  
    & 0.0511 & 0.0262
    & - & - 
    & - & - 
    & - & - \\ 

    \cmidrule(l){2-14}

    & \lgcn
    & 0.0108 & 0.0082
    & 0.0090 & 0.0062 
    & 0.0670 & 0.0336
    & - & - 
    & - & - 
    & - & - \\
 
    & \gfcf 
    & 0.0134 & 0.0112    
    & 0.0236 & 0.0175   
    & 0.0755 & 0.0379
    & - & - 
    & - & - 
    & - & - \\

    & \pgsp 
    & 0.0137 & 0.0112  
    & 0.0226 & 0.0168  
    & 0.0724 & 0.0535
    & - & - 
    & - & - 
    & - & - \\

    \midrule
    \multirow{5.5}{*}{\rotatebox[origin=c]{90}{\footnotesize Unified domain}}

    & \bpr
    & 0.0079 & 0.0047  
	& 0.0047 & 0.0019  
	& 0.0022 & 0.0095
    & 0.0011 & 0.0012 
	& 0.0013 & 0.0014  
	& 0.0014 & 0.0010 \\

    & \lightgcn 
	& 0.0320 & 0.0171   
	& \underline{0.0464} & 0.0316
	& 0.0255 & 0.0167
    & 0.0204 & 0.0473 
	& 0.0190 & 0.0351 
	& 0.0162 & 0.0183\\

    \cmidrule(l){2-14}

    & \lgcn 
	& 0.0282 & 0.0178   
	& 0.0366 & 0.0259   
	& 0.0684 & 0.0340
	& 0.0255 & 0.0543 
	& 0.0557 & 0.0804  
	& 0.0458 & 0.0447 \\
 
    & \gfcf 
    & 0.0334 & 0.0233  
	& 0.0424 & 0.0307   
	& 0.0684 & 0.0340
    & \textbf{0.0277} & \textbf{0.0611} 
	& 0.0609 & 0.0909
	& 0.0526 & 0.0504 \\

    & \pgsp 
	& \textbf{0.0360} & 0.0234
	& 0.0424 & 0.0307
	& 0.0684 & 0.0501
        & \underline{0.0277} & \underline{0.0611}
	& 0.0610 & 0.0909 
	& 0.0521 & 0.0498 \\

    \midrule
    \multirow{13.5}{*}{\rotatebox[origin=c]{90}{\footnotesize Cross-domain}}
    & \dcdcsr 
	& 0.0054 & 0.0044 
	& 0.0033 & 0.0027   
	& 0.0109 & 0.0047 
        & - & - 
        & - & - 
        & - & - \\

    & \conet
	& 0.0051 & 0.0035 
	& 0.0032 & 0.0023 
	& 0.0009 & 0.0005
        & - & - 
        & - & - 
        & - & - \\
    \cmidrule(l){2-14}

    & \emcdr
        & - & - 
        & - & - 
        & - & - 
	& 0.0087 & 0.0108
	& 0.0020 & 0.0039 
    & 0.0014 & 0.0012 \\

    & \sscdr
        & - & - 
        & - & - 
        & - & - 
	& 0.0022 & 0.0024
	& 0.0018 & 0.0014  
	& 0.0016 & 0.0013 \\
 
    & \ptupcdr
        & - & - 
        & - & - 
        & - & - 
	& 0.0088 & 0.0109
	& 0.0025 & 0.0047
	& 0.0015 & 0.0013 \\
    \cmidrule(l){2-14}

    & \unicdr
    & 0.0047 & 0.0033 
    & 0.0067 & 0.0061 
    & 0.0010 & 0.0012
    & 0.0209 & 0.0013
    & 0.0099 & 0.0041
    & 0.0049 & 0.0021 \\
    \cmidrule(l){2-14}


    & \proposedonef
    & 0.0276 & 0.0197 
	& 0.0436 & 0.0299 
	& 0.0764 & 0.0385
    & 0.0252 & 0.0550 
	& 0.0395 & 0.0640 
	& 0.0423 & 0.0404 \\
 
    & \proposedonev 
    & 0.0334 & \underline{0.0249}
	& 0.0236 & 0.0175 
	& 0.0670 & 0.0336
    & 0.0225 & 0.0467 
	& 0.0527 & 0.0752 
	& 0.0451 & 0.0443 \\

    & \proposedtwof
    & \underline{0.0349} & \textbf{0.0271}
	& \textbf{0.0479} & \textbf{0.0330} 
	& \underline{0.0965} & 0.0512
    & 0.0254 & 0.0556 
	& 0.0548 & 0.0906 
	& 0.0541 & 0.0636 \\
 
    & \proposedtwov 
    & 0.0283 & 0.0228 
	& \underline{0.0464} & \underline{0.0324}
	& \textbf{0.0977} & 0.0534
    & 0.0265 & 0.0587 
	& 0.0644 & 0.1065 
	& 0.0574 & 0.0672 \\

    & \proposedthreef
    & 0.0279 & 0.0213
	& 0.0356 & 0.0248
    & 0.0726 & \textbf{0.0538}
    & 0.0258 & 0.0474 
	& \underline{0.0660} & \underline{0.1126}
	& \underline{0.0651} & \underline{0.0793} \\
    
    &  \proposedthreev
    & 0.0179 & 0.0154 
	& 0.0262 & 0.0196 
	& 0.0727 & \underline{0.0537}
     & 0.0242 & 0.0579 
	& \textbf{0.0673} & \textbf{0.1136}
	& \textbf{0.0677} & \textbf{0.0861} \\

    \bottomrule
\end{tabular}
}
\end{table*}

\subsubsection{Baselines}
We consider different sets of baseline methods for intra-domain and inter-domain CDR scenarios, respectively.

For evaluation in \textbf{intra-domain CDR scenarios}, we execute basic CF methods using user-item interactions only in a target domain (\textit{single domain}), and those in both domains (\textit{unified domain}): 
\begin{itemize}
    \item \textbf{\gfcf}~\cite{shen2021powerful}, \textbf{\lgcn}~\cite{shen2021powerful}, \textbf{\pgsp}~\cite{liu2023personalized}: These are GSP-based methods specifically developed for CF. For \gfcf and \pgsp, a mixed-frequency filter that combines a linear filter and an ideal low-pass filter is initially used. To ensure a thorough comparison with our results, we conduct experiments using both the linear filter and the mixed-frequency filter, and report the superior performance.

    \item \textbf{\bpr}~\cite{rendle2012bpr}, \textbf{\lightgcn}~\cite{he2020lightgcn}: These are encoder-based methods that train user and item embeddings as well as a graph encoder by optimizing a pairwise ranking loss.
\end{itemize}
We also compare CDR methods for the intra-domain scenario:
\begin{itemize}\vspace{-\topsep}
    \item \textbf{\dcdcsr}~\cite{Zhu2018ADF} 
, \textbf{\conet}~\cite{hu2018conet}: These are encoder-based methods utilizing the dual-network architecture to transfer information through cross-connections by leveraging shared features between the source and target domains.
    \item \textbf{\unicdr}~\cite{cao2023towards}: This is the most recent and unified framework capable of both intra-domain and inter-domain recommendation by learning domain-specific feature extraction and shared feature representations.
\end{itemize}

For evaluation in \textbf{inter-domain CDR scenarios}, we consider basic CF methods using user-item interactions in a unified domain (\textit{unified domain}), where the union of both domains is regarded as a single target domain:
\textbf{\gfcf}~\cite{shen2021powerful}, \textbf{\lgcn}~\cite{shen2021powerful}, \textbf{\pgsp}~\cite{liu2023personalized}, \textbf{\bpr}~\cite{rendle2012bpr}, and \textbf{\lightgcn}~\cite{he2020lightgcn}.
We additionally evaluate CDR methods for the inter-domain scenario, including \textbf{\unicdr}~\cite{cao2023towards} which is applicable to intra-domain scenarios as well:

\begin{itemize}
    \item \textbf{\emcdr}~\cite{man2017cross},  \textbf{\sscdr}~\cite{kang2019semi}, \textbf{\ptupcdr}~\cite{zhu2022personalized}: These are encoder-based methods that adopt the embedding-and-mapping strategy to project cold-start users' embeddings from the source domain to the target domain.
\end{itemize}

For \proposed, we evaluate two configurations based on the hyperparameter $\alpha$, i.e., \textit{optimal} and \textit{fixed}, denoted as \textbf{$\proposed^*$} and \textbf{$\proposed^\dagger$}. 
$\proposed^*$ selects the $\alpha$ value that achieves the highest NDCG score on the validation set, while $\proposed^\dagger$ uses a fixed value of $\alpha=0.85$.


\subsection{Effectiveness in CDR Scenarios (RQ1\&2)}

\subsubsection{Performance for intra-domain recommendation (RQ1)}
The recommendation accuracy for target domain users (i.e., intra-domain) is reported in Table \ref{tbl:intra-inter-domain} (Left). 
\proposed shows significant performance improvements over single-domain methods, underscoring the value of leveraging source domain information, even with GSP-based CF methods.
Notably, on the Amazon Movie$\rightarrow$Music task with only 18\% overlapping users, it improves performance by nearly 30\% over the best baseline.
Additionally, \proposed also significantly outperform {the recent encoder-based CDR methods (i.e., \dcdcsr and \conet)}, attributable to the inherent challenges of CF with sparse dataset, where parametric encoders often struggle to be effectively optimized.
Compared to  baseline methods in the unified domain, \proposed consistently exhibits superior performance.
Notably, GSP-based CF methods using the unified domain graph do not always perform better than the ones using the single domain graph, implying their sensitivities to overlapping user ratio;
this suggests that incorporating source domain information by the cross-domain graph is more effective than simply merging two domains as a unified domain.
Among the three augmentation strategies within our \proposed framework, \proposedtwo especially demonstrates remarkable performance, which supports the importance of concentrating on overlapping users when {integrating information from external domains.}

\subsubsection{Performance for inter-domain recommendation (RQ2)} 
The recommendation accuracy for source domain users who have no interactions in target domain (i.e. cold-start users) is reported in Table \ref{tbl:intra-inter-domain} (Right). 
Both GSP-based CF methods that adopt the unified domain and cross-domain graph outperform encoder-based methods. 
Despite the absence of historical interactions in target domain, \proposed achieves good performance, indicating its effectiveness in addressing the cold-start problem.
As with intra-domain recommendation, encoder-based methods struggle to train mapping functions by solely relying on interactions without semantic features;
as the interaction matrix by itself lacks the comprehensive data necessary for effective learning in complex CDR scenarios. 
In this scenario, \proposedthree exhibits remarkable performance compared to \proposedone and \proposedtwo, which strongly indicates that augmentation with user-focused similarity is helpful to have a deep understanding of user behaviors. 
Given that \proposedthree takes substantial account of source domain information, it successfully captures personalized graph signals that enhance recommendation accuracy.

\subsection{Robustness to Overlapping User Ratio (RQ3)}
To assess the robustness of \proposed under varying levels of overlapping users between domains, we evaluate its performance along with CF baselines using a unified-domain graph on the \douban dataset for both intra-domain and inter-domain recommendation tasks.
As indicated in Table~\ref{table:task}, \douban exhibits a high overlap between source and target domains, which is rarely observed in real-world applications.
To simulate more realistic scenarios, we manually adjust the overlapping user ratio in the \douban Movie$\rightarrow$Music task, reducing it from nearly 100\% to 20\%  by randomly removing interactions of overlapping users in the source domain.

\begin{figure}[t]
    \centering
    \includegraphics[width=\linewidth]{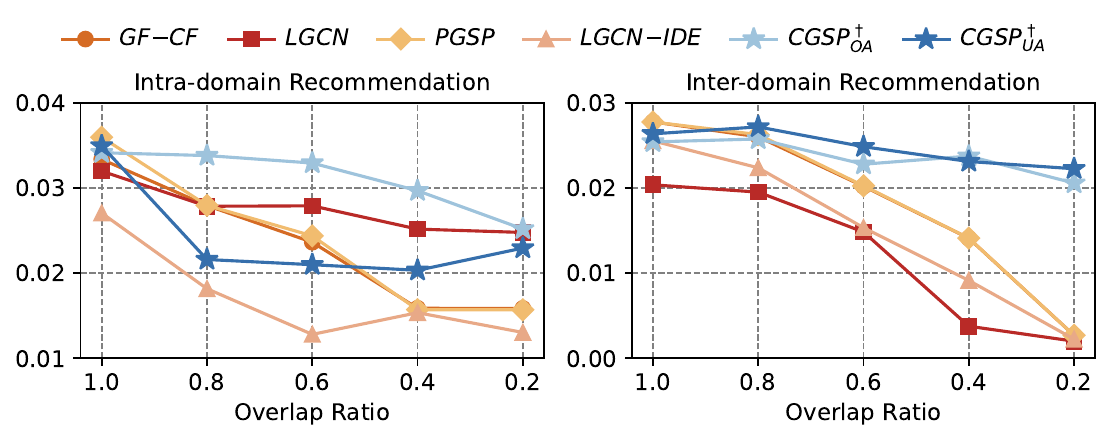}
    \vspace{-0.4cm}
    \caption{Performance (Recall@20) changes of \proposed and baseline methods in intra-domain (Left) and inter-domain (Right) CDR scenarios w.r.t. the ratio of overlapping users.}
    \label{fig:overlap-ratio}
\end{figure}

Figure~\ref{fig:overlap-ratio} shows how performance changes as the overlap ratio decreases.
While unified-domain baselines perform well under high overlap, where merging domains serves as simple data augmentation, their performance drops sharply under low-overlap conditions.
In contrast, \proposed remains stable with minimal variance across all overlap levels, significantly outperforming baselines when the overlap falls below 50\%.
Specifically, for intra-domain recommendation, \proposedtwof shows only a 26.21\% drop in performance from full to minimal overlap, whereas most baselines degrade by more than 50\%.
For the inter-domain recommendation task, \proposedthreef shows only 15.56\% drop,  in contrast to a severe 91.19\% performance loss observed in the worst-performing baseline.
These results demonstrate the robustness of \proposed under sparse overlap conditions, underscoring its effectiveness and practicality in real-world CDR scenarios where overlapping users are often limited.

\begin{figure}[t]
    \includegraphics[width=\linewidth]{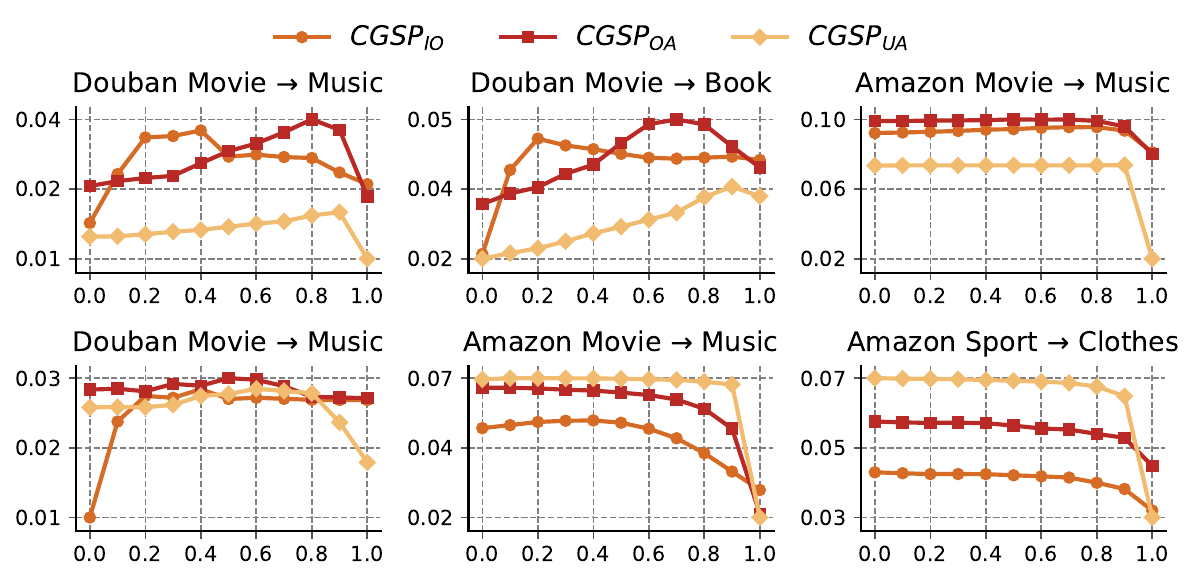}
    \vspace{-0.4cm}
    \caption{Performance (Recall@20) changes of \proposedone, \proposedtwo, and \proposedthree in intra-domain (Upper) and inter-domain (Lower) CDR scenarios w.r.t. hyperparameter $\alpha$.}
    \label{fig:alpha}
\end{figure}

\subsection{Impact of Source-Bridged Similarity (RQ4)}

We then investigate the impact of the hyperparameter $\alpha$ and its optimal value depending on the relationships between the two domains. 
Figure~\ref{fig:alpha} illustrates performance across different $\alpha$ values, ranging from 0.0 to 1.0, in both intra-domain (Upper) and inter-domain (Lower) recommendation scenarios. 
Note that $\alpha=0$ indicates exclusive reliance on target domain information, while $\alpha=1.0$ relies solely on source-bridged information.
We hypothesize that the semantic discrepancy between the source and target domains influences the utility of source-bridged information.

For intra-domain recommendation on the \douban Movie$\rightarrow$ Book task, the performance continues to increase as $\alpha$ increases, indicating that adding source-bridged information is beneficial;
On the other hand, in case of the Movie$\rightarrow$Music task for both \douban and \amazon, the performance initially improves with increasing $\alpha$ but declines sharply beyond a certain point, eventually performing worse than using only target domain information.
This suggests that Movie and Music domains are less semantically related than the Movie and Book.
Overall, optimal performance occurs when $\alpha$ lies between 0.7 and 0.9, where source-bridged information predominates over target-only. 
Conversely, extreme $\alpha$ values (0.0 or 1.0) consistently underperform, highlighting the benefit of a balanced integration.
For inter-domain recommendation, however, the performance does not improve significantly with increasing $\alpha$, and remains relatively strong even at $alpha$=0.
It implies that generating personalized input signals for each source-domain user is more crucial than the relative weighting of domain signals.


Meanwhile, across all scenarios in the \amazon dataset, performance shows little variation, likely due to the inherent sparsity in both the source and target domains.
We can conclude that the benefits derived from incorporating source domain are constrained under some conditions, such as low interaction density.

\subsection{Time Efficiency in CDR Scenarios (RQ5)}
We evaluate the efficiency of \proposed compared to encoder-based CDR baselines across both recommendation scenarios.
Table~\ref{tbl:time} reports the total execution time (for \proposed) and training time (for the baselines) required to attain their final outcomes presented in Table~\ref{tbl:intra-inter-domain}.
The results indicate that \proposed significantly reduces execution time compared to the baselines. 
The efficiency of \proposed is attributed to its avoidance of parameter optimization, making it both time-efficient and resource-friendly for real-world applications.

Typically, the matrix $ R \in \mathbb{R}^{m \times m} $, representing the graph with $ m $ entities, incurs a computational complexity of $ O(m^2)$ for matrix operations.
However, our framework significantly reduces the complexity in two key aspects.
First, the graph in our setting is highly sparse with the number of edges  $|E| \ll m^2 $.
This sparsity restricts operations to the non-zero entries, reducing the computational complexity to $ O(|E|) $. 
Second, \proposed employs a linear filter, which avoids additional computational overheads such as eigen-decomposition or iterative optimization, further simplifying the complexity.
This simplicity proves particularly beneficial in large-scale scenarios, as demonstrated by our experiments with the Amazon datasets.




\begin{table}[t]
\centering
\small
\caption{Total execution/training time of CDR methods in intra-domain (Upper) and inter-domain (Lower) recommendation scenarios, on \amazon Movie$\rightarrow$Music and \douban Movie$\rightarrow$Music. \lightgcn is executed on a unified domain. 
}
\label{tbl:time}
\footnotesize
\renewcommand{\arraystretch}{0.75}
\renewcommand{\tabcolsep}{1.2mm}
\centering
\resizebox{\linewidth}{!}{
\begin{tabular}{ccccccc}
\toprule
\textbf{Intra-} & \textbf{\proposedone} & \textbf{\proposedtwo} & \textbf{\proposedthree} & \textbf{\lightgcn} & \textbf{\dcdcsr} & \textbf{\unicdr} \\
\midrule
\amazon & 2m 36s & 3m 53s & 9m 59s & 55m 32s & 58m 18s & 205m 26s \\
\douban & 1m 56s & 2m 55s & 6m 34s & 71m 14s & 70m 20s & 377m 15s \\
\midrule
\textbf{Inter-} & \textbf{\proposedone} & \textbf{\proposedtwo} & \textbf{\proposedthree} & \textbf{\lightgcn} & \textbf{\sscdr} & \textbf{\unicdr} \\
\midrule
\amazon & 1m 4s & 1m 50s & 4m 28s & 26m 19s & 501m 45s & 247m 50s \\
\douban & 2m 3s & 2m 58s & 4m 7s & 33m 52s & 394m 37s & 206m 58s \\
\bottomrule
\end{tabular}
}
\end{table}


\begin{table}[t]
\centering
\caption{Performance of \proposedone using different graph filters, where $\phi$ denotes the linear filter’s proportion.}
\resizebox{0.97\linewidth}{!}{
\begin{tabular}{ll|ccccc}
\toprule
\multicolumn{2}{c|}{ $\phi$} & 1.00 & 0.75 & 0.50 & 0.25 & 0.00 \\
\midrule
\multirow{2}{*}{\textbf{Intra-}} & Recall@20 & \textbf{0.0764} & 0.0622 & 0.0505 & 0.0392 & 0.0313 \\
                              & NDCG@20   &\textbf{ 0.0385} & 0.0320 & 0.0260 & 0.0200 & 0.0165 \\
\midrule
\multirow{2}{*}{\textbf{Inter-}} & Recall@20 & \textbf{0.0395} & 0.0347 & 0.0290 & 0.0224 & 0.0172 \\
& NDCG@20   & \textbf{0.0640} & 0.0579 & 0.0497 & 0.0420 & 0.0372 \\
\bottomrule
\end{tabular}
}
\label{tbl:filter}
\end{table}

\subsection{Effect of Various Graph Filters}

We explore the potential benefits of applying various types of filters within our framework, even though we mainly focus on linear filters in this work. 
We analyze items-only cross-domain similarity graph $\mathbf{G}_{\textsc{io}}$ from \proposedone 
with two different graph filtering methods: a linear filter and an ideal low-pass filter. 
We mix two filters by controlling the proportion of the linear filter by $\phi$.
In Table~\ref{tbl:filter}, applying only the linear filter ($\phi=1$) shows the best performance;
this is because our CDR dataset has a high sparsity, which makes it challenging for the ideal low-pass filter to capture user preferences from global signals. 
However, there is still much room for further improvement in performance by utilizing various graph filters.

\section{Related Work}
\label{sec:relatedwork}
\newcommand{\cmt}[1]{\textcolor{red}{#1}}


\smallsection{Cross domain recommendation}
The existing CDR studies can be categorized according to the target CDR scenarios: intra-domain and inter-domain recommendation.
For intra-domain recommendation~\cite{li2020ddtcdr, cao2022disencdr, liu2020cross, zhao2020catn, yuan2019darec, yang2024not, li2023one, xu2023cdaml, hu2018conet, Zhu2018ADF, chen2018deep}, many studies have mitigated the data sparsity in the target domain by leveraging source domain information.
DDSAN~\cite{chen2018deep} proposes a new regularization technique, which aligns model parameters across domains, to promote domain invariance.
CoNet~\cite{hu2018conet} enables feature learning with cross-connections in its neural network.
CDAML~\cite{xu2023cdaml} utilizes clustering alongside meta-learning for domain adaptation and
CAT-ART~\cite{li2023one} generates global user embeddings using a contrastive autoencoder.
UniCDR~\cite{cao2023towards} employs domain-specific and domain-shared embeddings with aggregation schemes, handling both~two~CDR~scenarios. \
The inter-domain recommendation~\cite{liu2023contrastive, liu2024user, lu2024amt, zhao2023cross, zhu2019dtcdr, zhu2020graphical, sahu2020knowledge, man2017cross, cao2022cross, zhu2022personalized} aims to provide recommendations to cold-start users who have not interacted with the target domain, by leveraging 
source domain information.
Many studies have focused on transfer learning based on shared user features or mapping functions.
EMCDR~\cite{man2017cross} uses latent representations and mapping functions for knowledge transfer, and SSCDR~\cite{kang2019semi} further proposes a semi-supervised learning method to effectively train the mapping function.
CDRIB \cite{cao2022cross} introduces new regularizers based on the information bottleneck principle to build user-item correlations across domains.
PTUPCDR~\cite{zhu2022personalized} proposes a meta-network to generate personalized mapping functions.
UniCDR \cite{cao2023towards} can also be applied to recommend inter-domain recommendations using domain-shared embeddings.
Though effective, existing CDR methods rely heavily on overlapping users as a bridge for knowledge transfer during model training. Additionally, they lack the flexibility to adjust the impact of the source domain; it makes them less effective when two domains have large intrinsic discrepancies in user behavior.

\smallsection{GSP-based Collaborative Filtering}
Recently, graph signal processing has gained significant attention as a promising direction for collaborative filtering~\cite{shen2021powerful, liu2022parameter, liu2023personalized, xia2024hierarchical, xia2022fire, liu2022collaborative, choi2023bspm, hong2024svdae, peng2024powerful, park2025criteria}
, as its primary concept aligns with the goal of collaborative filtering
(explained in Section~\ref{sec:preliminary}).
\gfcf~\cite{shen2021powerful} demonstrates that graph filtering methods are closely connected to existing CF methods.  
Motivated by the linear filter in \lightgcn~\cite{he2020lightgcn}, they suggested \lgcn and \gfcf, a unified graph convolution framework using a mixed-frequency graph filter combining linear and ideal low-pass filter.
FIRE~\cite{xia2022fire} extends this approach for real-time recommendation, addressing dynamic interactions and the cold-start problem.
PGSP~\cite{liu2023personalized} further develops this concept by incorporating an augmented similarity graph with personalized graph signals.
HiGSP~\cite{xia2024hierarchical} advanced the framework with cluster-wise and globally-aware filters.
While GSP has proven effective in single-domain recommendations, its effectiveness in CDR remains unexplored in the literature.


\section{Conclusion}
\label{sec:conclusion}

This paper presents a novel and unified CDR framework, termed \proposed, based on GSP to effectively capture both intra-domain and inter-domain relationships through a cross-domain similarity graph.
In particular, we introduce the hyperparameter $alpha$ hat adaptively controls the influence of source-domain information, addressing the sensitivity of CDR tasks to domain discrepancies.
\proposed demonstrates substantial improvements in recommendation performance and time efficiency coover strong baselines, particularly under low overlapping-user conditions.
Our research paves the way for broader applications of GSP in real-world scenarios of cross-domain recommendation.




\bibliographystyle{ACM-Reference-Format}
\bibliography{bibliography}


\end{document}